\documentclass[twocolumn,showpacs]{revtex4}

\usepackage{amsmath}
\usepackage{amsfonts}
\usepackage{amssymb}
\usepackage{bm}
\usepackage[english]{babel}

\begin{document}

\title{Euler Chern Simons Gravity from Lovelock Born Infeld Gravity}

\author{Fernando Izaurieta}
\email{fizaurie@udec.cl}
\author{Eduardo Rodr\'{\i}guez}
\email{edurodriguez@udec.cl}
\author{Patricio Salgado}
\email{pasalgad@udec.cl}

\affiliation{Departamento de F\'{\i}sica, Universidad de
Concepci\'on, Casilla 160-C, Concepci\'on, Chile}
\date{\today}

\begin{abstract}
In the context of a gauge theoretical formulation, higher
dimensional gravity invariant under the $AdS$ group is
dimensionally reduced to Euler-Chern-Simons gravity. The
dimensional reduction procedure of Grignani-Nardelli [Phys. Lett.
B \textbf{300}, 38 (1993)] is generalized so as to permit reducing
$D$-dimensional Lanczos Lovelock gravity to $d=D-1$ dimensions.
\end{abstract}

\maketitle

\section{Introduction}

Odd-dimensional gravity may be cast as a gauge theory for the
(A)dS groups \cite{Cha89}. The lagrangian is the
Euler-Chern-Simons form in $D=2n-1$ dimensions \cite{Ban94,Zan00}
\[
S_{\text{CS}}^{\left( D\right) }=\int \sum_{p=0}^{\left[ D/2\right] }\alpha
_p^{\left( D\right) }{\cal L}_p^{\left( D\right) },
\]
where
\[
{\cal L}_p^{\left( D\right) }=\varepsilon _{A_1\cdots A_D}R^{A_1A_2}\cdots
R^{A_{2p-1}A_{2p}}e^{A_{2p+1}}\cdots e^{A_D},
\]
\[
\alpha _p^{\left( D\right) }=\kappa \frac{l^{-\left( D-2p-1\right) }}{\left(
D-2p\right) }%
%TCIMACRO{\binom{\frac{D-1}2}p}%
%BeginExpansion
{\frac{D-1}2 \choose p}%
%EndExpansion
,
\]
whose exterior derivative is Euler's topological invariant in $2n$
dimensions. The constants $\kappa $ and $l$ are related to Newton's constant
$G$ and to the cosmological constant $\Lambda $ through $1/\kappa =2\left(
D-2\right) !\Omega _{D-2}G$ (where $\Omega _{D-2}$ is the area of the $%
\left( D-2\right) $ unit sphere) and $\Lambda =\pm \left( D-1\right) \left(
D-2\right) /2l^2$.

The Chern-Simons lagrangian remains invariant under local Lorentz rotations
in tangent space $\delta e^A=\lambda _{\;B}^Ae^B$, $\delta \omega
^{AB}=-D\lambda ^{AB},$ and changes by a total derivative under an
infinitesimal (A)dS boost $\delta e^A=-D\lambda ^A$, $\delta \omega ^{AB}=%
\frac 1{\ell ^2}\left( \lambda ^Ae^B-\lambda ^Be^A\right) $ . With
appropriate boundary conditions, this means that the action is left
invariant by the (A)dS gauge transformations. Furthermore, the vielbein and
the spin connection correspond to the gauge fields associated with (A)dS
boosts and Lorentz rotations, respectively. Thus, odd-dimensional
Chern-Simons gravity is a good gauge theory for the (A)dS group, but its
usefulness is limited to odd dimensions. This is related to the fact that no
topological invariants have been found in even dimensions, and therefore the
derivative of an even-dimensional lagrangian cannot be made equal to any of
them \cite{Zan00}.

The requirement that the equations of motion fully determine the
dynamics for as many components of the independent fields as
possible may also be used in the even-dimensional case, leading to
the so-called Lovelock-Born-Infeld action \cite{Ban94,Zan00}
\begin{equation}
S_{\text{BI}}^{\left( 2n\right) }=\frac{\kappa l^2}{2n}\int \varepsilon
_{A_1\cdots A_{2n}}\bar{R}^{A_1A_2}\cdots \bar{R}^{A_{2n-1}A_{2n}},
\label{SBI2n}
\end{equation}
where $\bar{R}^{AB}\equiv R^{AB}+\frac 1{l^2}e^Ae^B$ and $l$ is a length.
For $D=2n=4$, (\ref{SBI2n}) reduces to the EH action with a cosmological
constant $\Lambda =\pm 3/l^2$ plus Euler's topological invariant with a
fixed weight factor.

When one considers the spin connection $\omega ^{AB}$ and the vielbein $e^A$
as components of a connection for the (A)dS group, one finds that the action
(\ref{SBI2n}) is invariant under local Lorentz rotations while, under
infinitesimal (A)dS boosts, it changes by
\[
\delta S_{\text{BI}}^{\left( 2n\right) }=-\kappa \int \varepsilon
_{A_1\cdots A_{2n}}\bar{R}^{A_1A_2}\cdots \bar{R}%
^{A_{2n-3}A_{2n-2}}T^{A_{2n-1}}\lambda ^{A_{2n}},
\]
where $\lambda ^A$ is the infinitesimal parameter of the transformation. It
simply makes no sense to use the equations of motion associated with the
action (\ref{SBI2n}) to enforce the invariance; {\em any} action is on-shell
invariant under {\em any} infinitesimal transformation, just by the
definition of the equations of motion. On the other hand, one could try to
set the torsion equal to zero by fiat, and impose $T^A=0$ as an off-shell
identity. This is unsatisfactory in the sense that, in the (A)dS gauge
picture, torsion and curvature stand on a similar footing as fields
strengths of the connection whose components are $e^A$ and $\omega ^{AB}$.
It would seem rather odd to have some components of the field strength set
arbitrarily to zero while the others remain untouched. The gauge
interpretation of even-dimensional Lovelock-Born-Infeld gravity is thus
spoiled by the lack of invariance of the action (\ref{SBI2n}) under
infinitesimal (A)dS boosts.

A truly (A)dS-invariant action for even as well as for odd dimensions was
constructed in ref. \cite{Sal03} using the Stelle-West formalism \cite{Ste80}%
\ for non-linear gauge theories. The action is
\[
S_{\text{SW}}^{\left( D\right) }=\int \sum_{p=0}^{\left[ D/2\right] }\alpha
_p\varepsilon _{A_1\cdots A_d}{\cal R}^{A_1A_2}\cdots {\cal R}%
^{A_{2p-1}A_{2p}}V^{A_{2p+1}}\cdots V^{A_D},
\]
where
\[
{\cal R}^{AB}=dW^{AB}+W_{\;C}^AW^{CB},
\]
\[
V^A=\Omega _{\;B}^A\left( \cosh z\right) e^B+\Omega _{\;B}^A\left( \frac{%
\sinh z}z\right) D_\omega \zeta ^B,
\]
\[
W^{AB}=\omega ^{AB}+\frac \sigma {l^2}\left[ \left( \frac{\sinh z}z\right)
e^C+\left( \frac{\cosh z-1}{z^2}\right) D_\omega \zeta ^C\right] \cdot
\]
\[
\cdot \left( \zeta ^A\delta _C^B-\zeta ^B\delta _C^A\right) ,
\]
with
\begin{equation}
\Omega _{\;B}^A\left( u\right) \equiv u\delta _B^A+\left( 1-u\right) \frac{%
\zeta ^A\zeta _B}{\zeta ^2}.  \label{Om}
\end{equation}
Here $\zeta ^A$ corresponds to the so-called AdS coordinate, which
parametrizes the coset space $SO\left( D+1\right) /SO\left( D\right) $, and $%
z=\zeta /l$.

The method devised in ref. \cite{Sal03} allows for the even-dimensional
action to become (A)dS gauge invariant. The same construction can be applied
in odd dimensions, where its only outcome is the addition of a boundary term
to the Chern-Simons action.

$(2n-1)$-dimensional gravity has attracted a growing attention in
recent years, both as a good theoretical laboratory for a possible
quantum theory of gravity and as a limit of the so called M
theory. In this context it is then interesting to establish a
clear link between $D=2n$ y $D=2n-1$ gravities by a dimensional
reduction. This is the aim of the present letter and it is
achieved in the framework of a gauge theoretical formulation of
both theories. In fact, as is shown in \cite{Zan00,Sal03}, gravity
in $2n-1$ and $2n$ dimensions can be formulated as a gauge theory
of the $AdS $ group. In $2n-1$ dimensions this formulation is
especially attractive as the Lanczos-Lovelock action becomes the
Chern-Simons term of the $AdS$ group. Such a Chern-Simons action
with the correct $AdS$ gauge transformations can then be derived
by dimensionally reducing the $2n$ dimensional Lanczos-Lovelock
action in its gauge theoretical formulation.

In \cite{Gri93} was proved, in the context of a Poincar\'{e} gauge
theoretical formulation, that pure gravity in $3+1$ -dimensions can be
dimensionally reduced to gravity in $2+1$ dimensions. However, the mechanism
of Grignani-Nardelli is not applicable in the context of an AdS gauge
theoretical formulation. One of the goals of this paper is to find a
generalization of the procedure of Grignani-Nardelli that permits, in the
context of an AdS gauge theoretical formulation, to reduce $D$-dimensional
LL gravity to $d\equiv D-1$ dimensions.

\section{Grignani Nardelli Procedure and AdS invariance}

Latin letters from the beginning of the alphabet will be used for tangent
space indices; $A,B,C=1,2,\ldots ,D$, and $a,b,c=1,2,\ldots ,d$. Greek
letters and latin letters from the middle of the alphabet will denote
space-time indices; $\lambda ,\mu ,\nu =1,2,\ldots ,D$, and $%
i,j,k=1,2,\ldots ,d$. Fields belonging to $d=D-1$ dimensions will be
distinguished by underlining, as in \underline{$\omega $}$^{ab}$. Exterior
derivatives will be denoted as $d=dx^\mu \partial _\mu $ and $\underline{d}%
=dx^i\partial _i$. First, we consider the dimensional reduction from $(3+1)$
to $(2+1)$ dimensions. The Lovelock Born Infeld lagrangian in $D=4$ is given
by
\[
L_{\text{BI}}^{\left( 4\right) }=\frac \kappa 4\varepsilon _{ABCD}\left(
{\cal R}^{AB}+\frac 1{l^2}{\bm V}^A{\bm V}^B\right) \left( {\cal R}^{CD}+%
\frac 1{l^2}{\bm V}^C{\bm V}^D\right) ,
\]
where
\[
{\cal R}^{AB}\equiv dW^{AB}+W_{\;C}^AW^{CB}
\]
is the curvature tensor. This lagrangian can be written in the form

\[
L_{\text{BI}}^{\left( 4\right) }=\kappa \varepsilon _{abc}\left( {\cal R}%
^{ab}+\frac 1{l^2}{\bm V}^a{\bm V}^b\right) \left( {\cal R}^{c4}+\frac 1{l^2}%
{\bm V}^c{\bm V}^4\right) ,
\]
where we have used $\varepsilon _{abc4}=\varepsilon _{abc}$.

\textbf{Method I:} The table A of \cite{Gri93} can be written as:
\[
e^A=\left( {\bm e}^a{\bm ,e}^{2n}\right) =\left(
e^a,dx^{2n}\right) ,
\]
\[
\omega ^{AB}=\left[
\begin{array}{cc}
{\bm \omega }^{ab} & {\bm \omega }^{a,2n} \\
{\bm \omega }^{2n,b} & {\bm \omega }^{2n,2n}%
\end{array}
\right] =\left[
\begin{array}{cc}
{\bm \omega }^{ab} & 0 \\
0 & 0%
\end{array}
\right] ,
\]
\[
\zeta ^A=\left( {\bm \xi }^a{\bm ,\zeta }^{2n}\right) =\left(
\zeta ^a,0\right) .
\]
This means
\[
{\bm V}^a=V^a,
\]
\[
{\bm V}^4=\left( \cosh z\right) dx^4,
\]
\[
{\cal R}^{ab}=R^{ab},
\]
\[
{\cal R}^{a4}=-m^2D_W\left[ \left( \frac{\sinh z}z\right) \zeta ^a\right]
dx^4.
\]

By substituting these results in $L_{\text{BI}}^{\left( 4\right) }$ one gets
\[
L_{\text{red}}^{\left( 3\right) }=\frac \kappa \ell \varepsilon
_{abc}\{\left( \cosh z\right) \left( R^{ab}+\frac 1{\ell ^2}V^aV^b\right)
V^c
\]
\[
-\left( R^{ab}+\frac 1{\ell ^2}V^aV^b\right) D_W\left[ \left( \frac{\sinh z}z%
\right) \zeta ^c\right] \}.
\]
Since
\[
\varepsilon _{abc}V^aV^bD_WA^c=d\left( \varepsilon _{abc}V^aV^bA^c\right)
-2\varepsilon _{abc}{\cal T}^aV^bA^c,
\]
we have
\[
L_{\text{red}}^{\left( 3\right) }=\frac \kappa \ell \left( \cosh z\right)
\varepsilon _{abc}\left( R^{ab}+\frac 1{\ell ^2}V^aV^b\right) V^c
\]
\[
+\frac{2\kappa }{\ell ^3}\left( \frac{\sinh z}z\right) \varepsilon _{abc}%
{\cal T}^aV^b\zeta ^c+\text{surface term}
\]
which is very different of the Chern-Simons Lagrangian in $2+1$
dimensions.

\textbf{Method II:} The table B of \cite{Gri93} can be written as
\[
e^A=\left( {\bm e}^a{\bm ,e}^{2n}\right) =\left(
e^a,dx^{2n}\right) ,
\]
\[
\omega ^{AB}=\left[
\begin{array}{cc}
{\bm \omega }^{ab} & {\bm \omega }^{a,2n} \\
{\bm \omega }^{2n,b} & {\bm \omega }^{2n,2n}%
\end{array}
\right] =\left[
\begin{array}{cc}
\omega ^{ab} & \frac 1\gamma V^a \\
-\frac 1\gamma V^b & 0%
\end{array}
\right] ,
\]
\[
\zeta ^A=\left( {\bm \zeta }^a{\bm ,\zeta }^{2n}\right) =\left(
0,\gamma \right) .
\]

This means

\[
{\bm V}^a=\left( \frac{\sinh z}z\right) V^a,
\]
\[
{\bm V}^4=dx^4,
\]
\[
{\cal R}^{ab}=R^{ab}-\frac 1{\gamma ^2}\left( \cosh ^2\hat{z}\right) V^aV^b,
\]
\[
{\cal R}^{a4}=\frac 1\gamma \left( \cosh z\right) D_\omega V^a,
\]
with $z=m\gamma $.

By substituting these results in $L_{\text{BI}}^{\left( 4\right) }$ one gets
\[
L_{\text{red}}^{\left( 3\right) }=\frac \kappa l\left( \frac{\sinh z}z%
\right) \varepsilon _{abc}\left( R^{ab}-\frac 1{\gamma ^2}V^aV^b\right) V^c,
\]
which it is again different from the Chern-Simons Lagrangian in $2+1$
dimensions. Similar results are obtained for higher dimensions.

\section{Generalization}

We now consider a generalization of the mechanism of Grignani Nardelli. The
basic idea is the following. Let $S_D$ be an action functional defined as
the integral of a lagrangian $D$-form ${\cal L}_D$ over a $D$-dimensional
manifold $M_D$. Let us explicitly perform one integration out of the $D$. We
are then left with the integral of a $d$-form over a $d$-dimensional
manifold $M_d$. Relabel the fields in this integral in a convenient way and
call the $d$-form ${\cal L}_d$. Its integral over $M_d$ then becomes what we
call $S_d$. Therefore, in general,
\begin{equation}
S_D=\int_{M_D}{\cal L}_D=\int_{M_d}{\cal L}_d=S_d,
\end{equation}
where
\begin{equation}
{\cal L}_d=\int_{\text{1 dim}}{\cal L}_D.
\end{equation}

Consider a surface $\sigma \left( x\right) =$ const and a vector field $n$
not lying on the surface; i.e., $I_n\left( d\sigma \right) \neq 0,$ where $%
I_\xi $ is the contraction operator. It will prove convenient to normalize
the vector $n$ to make it fulfill the condition $I_n\left( d\sigma \right)
=1 $.

A $p$-form field $\psi $ living in $D$ dimensions can always be decomposed
according to
\begin{equation}
\psi =\hat{\psi}+\check{\psi},  \label{c=c+c}
\end{equation}
where we have defined $\hat{\psi}\equiv d\sigma I_n\psi ,$ $\check{\psi}%
\equiv I_n\left( d\sigma \psi \right) .$ The two fields $\hat{\psi}$ and $%
\check{\psi}$ together carry the same information as the original field $%
\psi $, as can be seen from (\ref{c=c+c}). The $\check{\psi}$-component of $%
\psi $ lies entirely on the surface $\sigma \left( x\right) =$ const, while $%
\hat{\psi}$ retains that part of $\psi $ which goes in the direction of $n$.

When this decomposition is applied to the $D$-form lagrangian ${\cal L}_D$,
one finds $\stackrel{\wedge }{{\cal L}}_D=d\sigma I_n{\cal L}_D,$ $\stackrel{%
\vee }{{\cal L}}_D=0.$ This means that it is possible to integrate ${\cal L}%
_D=\stackrel{\wedge }{{\cal L}}_D$ over $\sigma $ to obtain the $d$%
-dimensional lagrangian ${\cal L}_d$ as
\begin{equation}
{\cal L}_d=\int_\sigma d\sigma I_n{\cal L}_D.
\end{equation}
The action is written as
\begin{equation}
S_D=\int_{M_d}{\cal L}_d,
\end{equation}
where $M_d$ is a $d$-dimensional manifold that belongs to the equivalence
class of manifolds induced by $\sigma $ (an example is shown below). It is
perhaps interesting to note that so far there is no need to assume that the
integrated direction has any especial feature such as being compact or
extremely curved; the reduction procedure remains well-defined whether we
make these assumptions or not.

Now we consider the simple case $\sigma \left( x\right) =x^D$; that is, we
deal with slices of constant $x^D$ across $D$-dimensional space-time. A
natural choice for $n$ is thus $n=\partial _D\equiv \partial /\partial x^D$,
which satisfies $I_nd\sigma =I_{\partial _D}dx^D=1$. With these choices, the
$d$-form lagrangian ${\cal L}_d$ may be written as
\begin{equation}
\left( {\cal L}_d\right) _{i_1\cdots i_d}=\int_{x^D}\left( {\cal L}_D\right)
_{i_1\cdots i_dD}dx^D,
\end{equation}
and $M_d$ is simply any $x^D=$ const manifold. For {\em all} of them, the
integral
\begin{equation}
S_d=\int_{M_d}{\cal L}_d  \label{S=8L}
\end{equation}
has the same value. However, the relabeling of the fields that remains to be
done in (\ref{S=8L}) may be more natural on one specific surface.

There is much freedom in the way this field-relabeling process is performed,
as the only strong constraints come from symmetries. In general, the
original fields in ${\cal L}_D$ transform locally under a group $G$ over $%
M_D $, leaving ${\cal L}_D$ invariant. On the other hand, the relabeled
fields that enter ${\cal L}_d$ must transform locally under a group $%
G^{\prime }$ over $M_d$, leaving ${\cal L}_d$ invariant. However, the
lagrangian ${\cal L}_d$ is still invariant under the $G$ group, which gets
realized now in a different way. Thus, the relabeling process must be
carried out in such a way that this requirement is satisfied. A good example
is provided by the spin connection $\omega ^{AB}$. Under a local,
infinitesimal Lorentz transformation $\Lambda =1+\frac 12\lambda ^{AB}{\bm J}%
_{AB}$ defined over $M_D$, it changes by $\delta \omega ^{AB}=-D\lambda
^{AB}.$ Our question now is, what components of this $SO\left( D\right) $
connection may be relabeled as the $SO\left( d\right) $ connection
\underline{$\omega $}$^{ab}$? To find the answer, perform an $M_d$-local $%
SO\left( d\right) $ transformation on $\omega ^{AB}$, i.e., demand that the $%
SO\left( D\right) $ infinitesimal parameters $\lambda ^{AB}$ satisfy the
conditions $\partial _D\lambda ^{AB}=0,$ $\lambda ^{A,D}=0$ . These
conditions turn the remaining $\lambda ^{ab}$ into the right parameters for
a $SO\left( d\right) $ infinitesimal transformation. It is straightforward
to show that, when this is the case, we have $\delta \check{\omega}^{ab}=-%
\check{D}\lambda ^{ab},$ $\delta \hat{\omega}^{ab}=\lambda _{\;c}^a\hat{%
\omega}^{cb}+\lambda _{\;c}^b\hat{\omega}^{ac},$ where $\check{D}$ is the
exterior covariant derivative in the connection $\check{\omega}^{ab}$. These
last eqs. express that $\check{\omega}^{ab}$ transforms as a $SO\left(
d\right) $-connection, while $\hat{\omega}^{ab}$ behaves as a $SO\left(
d\right) $-tensor. Therefore, an identification such as
\begin{equation}
\check{\omega}^{ab}\rightarrow \underline{\omega }^{ab}+\left[ SO\left(
d\right) \text{-tensor}\right] ^{ab}  \label{ww}
\end{equation}
seems quite reasonable. In general, one simple way to respect the relevant
symmetries is to identify the components of a $D$-dimensional field with
those of the corresponding $d$-dimensional one.

Here we shall perform this kind of identifications in their simplest
possible form. Many components of the fields will be frozen to zero; this
corresponds to our early assertion that we are only interested (by now) in
showing the possibility of getting $d$-dimensional gravity from its
higher-dimensional version. This is no longer true, of course, when we face
the full dimensional reduction procedure, for in this case freezing some
components of the fields yields a reduced gauge group as well.

\section{The Lanczos Lovelock action}

We shall start our dimensional reduction process with the $D$-dimensional LL
action
\begin{equation}
S_D=\int \sum_{p=0}^{\left[ \frac D2\right] }\alpha _p\varepsilon
_{A_1\cdots A_d}{\cal R}^{A_1A_2}\cdots {\cal R}%
^{A_{2p-1}A_{2p}}V^{A_{2p+1}}\cdots V^{A_d}.  \label{SD}
\end{equation}

A more suitable version of action (\ref{SD}) is
\[
S_D=\int_{M_D}\sum_{p=0}^{\left[ \frac{D-1}2\right] }\left( D-2p\right)
\alpha _p\varepsilon _{a_1\cdot \cdot \cdot a_d}{\cal R}^{a_1a_2}\cdot \cdot
{\cal R}^{a_{2p-1}a_{2p}}V^{a_{2p+1}}\cdot \cdot
\]
\[
\cdot \cdot \cdot V^{a_d}V^D+2\left( -1\right) ^D\int_{M_D}\sum_{p=1}^{\left[
D/2\right] }p\alpha _p\varepsilon _{a_1\cdots a_d}{\cal R}^{a_1a_2}\cdots
\]
\begin{equation}
\cdot \cdot \cdot {\cal R}^{a_{2p-3}a_{2p-2}}{\cal R}^{a_{2p-1},D}V^{a_{2p}}%
\cdots V^{a_d}.  \label{Sd2}
\end{equation}

The first term in (\ref{Sd2}) shows that it is {\em always} possible to
obtain a LL action in $d$ dimensions starting from a LL action defined in $%
D=d+1$.

The coefficients $\alpha _p$ in (\ref{SD}) are selected according to the
criterion that the equations of motion fully determine the dynamics for as
many components of the independent fields as possible. This analysis leads
to \cite{Sal03}
\begin{equation}
\alpha _p=\left\{
\begin{array}{cc}
\frac{\kappa _D}{D-2p}l^{-\left( D-2p-1\right) }%
%TCIMACRO{\binom{\frac{D-1}2}p }%
%BeginExpansion
{\frac{D-1}2 \choose p}
%EndExpansion
& \text{when }D\text{ is odd} \\
\kappa _D\frac{\kappa _D}Dl^{-\left( D-2p-1\right) }%
%TCIMACRO{\binom{\frac D2}p }%
%BeginExpansion
{\frac D2 \choose p}
%EndExpansion
& \text{when }D\text{ is even.}%
\end{array}
\right.  \label{alfa}
\end{equation}

A well-defined dynamics in $D$ dimensions leads to a well-defined dynamics
in $d$ dimensions; however, we shall additionally demand that the purely
gravitational terms in the reduced action produce well-defined dynamics as
well. This means that the coefficients $\alpha _p$ must be reduced
accordingly; that is, the coefficients in the reduced gravitational
lagrangian must correspond to $\alpha _p^{\left( d\right) }$ as given in (%
\ref{alfa}), with $D\rightarrow d$.

We consider the dimensional reduction from $D=$even to $d=D-1=$odd. First we
note that the action (\ref{SD}) includes Euler's topological invariant for $%
p=D/2$, and we may write it as
\[
S_D=\int_{M_D}\sum_{p=0}^{\left[ d/2\right] }\alpha _p^{\left( D\right)
}\varepsilon _{A_1\cdots A_D}{\cal R}^{A_1A_2}\cdots {\cal R}%
^{A_{2p-1}A_{2p}}V^{A_{2p+1}}\cdots
\]
\[
\cdot \cdot \cdot V^{A_D}+\alpha _{D/2}^{\left( D\right)
}\int_{M_D}\varepsilon _{A_1\cdots A_D}{\cal R}^{A_1A_2}\cdots {\cal R}%
^{A_{D-1}A_D}.
\]
This action is decomposed as
\[
S_D=\int_{M_D}\sum_{p=0}^{\left[ d/2\right] }\left( D-2p\right) \alpha
_p^{\left( D\right) }\varepsilon _{a_1\cdot \cdot \cdot a_d}{\cal R}%
^{a_1a_2}\cdot \cdot \cdot
\]
\[
\cdot \cdot {\cal R}^{a_{2p-1}a_{2p}}V^{a_{2p+1}}\cdot \cdot
V^{a_d}V^D+2\int_{M_D}\sum_{p=1}^{\left[ d/2\right] }p\alpha _p^{\left(
D\right) }\varepsilon _{a_1\cdot \cdot \cdot a_d}\cdot
\]
\[
\cdot {\cal R}^{a_1a_2}\cdot \cdot \cdot {\cal R}^{a_{2p-3}a_{2p-2}}{\cal R}%
^{a_{2p-1}D}V^{a_{2p}}\cdots V^{a_d}
\]
\begin{equation}
+\alpha _{D/2}^{\left( D\right) }\int_{M_D}\varepsilon _{A_1\cdots A_D}{\cal %
R}^{A_1A_2}\cdots {\cal R}^{A_{D-1}A_D}.  \label{c79}
\end{equation}
In this case the relation between the $\alpha _p$ coefficients in $D$ and $d$
dimensions is given by [cf. eq. (\ref{alfa})]
\begin{equation}
\frac 1l\frac{\kappa _D}{\kappa _d}\left( d-2p\right) \alpha _p^{\left(
d\right) }=\left( D-2p\right) \alpha _p^{\left( D\right) }.  \label{c87}
\end{equation}
Plugging (\ref{c87}) into (\ref{c79}), we are led to
\begin{equation}
S_D=S_D^{\left( \text{G}1\right) }+S_D^{\left( \text{I}\right) }+S_D^{\left(
\text{G}2\right) },
\end{equation}
where
\[
S_D^{\left( \text{G1}\right) }=\frac 1l\frac{\kappa _D}{\kappa _d}%
\int_{M_D}\sum_{p=0}^{\left[ d/2\right] }\alpha _p^{\left( d\right) }\left(
d-2p\right) \varepsilon _{a_1\cdots a_d}{\cal R}^{a_1a_2}\cdots
\]
\begin{equation}
\cdot \cdot \cdot {\cal R}^{a_{2p-1}a_{2p}}V^{a_{2p+1}}\cdots V^{a_d}V^D,
\label{SG1}
\end{equation}
\[
S_D^{\left( \text{I}\right) }=2\int_{M_D}\sum_{p=1}^{\left[ d/2\right]
}p\alpha _p^{\left( D\right) }\varepsilon _{a_1\cdots a_d}{\cal R}%
^{a_1a_2}\cdots
\]
\begin{equation}
\cdot \cdot \cdot {\cal R}^{a_{2p-3}a_{2p-2}}{\cal R}^{a_{2p-1}D}V^{a_{2p}}%
\cdots V^{a_d},  \label{SI}
\end{equation}
\begin{equation}
S_D^{\left( \text{G2}\right) }=\frac{\kappa _Dl}D\int_{M_D}\varepsilon
_{A_1\cdot \cdot \cdot A_D}{\cal R}^{A_1A_2}\cdot \cdot {\cal R}%
^{A_{D-1}A_D}.  \label{SG2}
\end{equation}

Now we show that it is possible to obtain an action for $d=D-1$ dimensional
gravity from the corresponding action in $D$ dimensions. Any identification
in the spirit of (\ref{ww}) will do the job; for example,
\begin{equation}
\check{V}^a\rightarrow \underline{V}^a,\hat{V}^D\rightarrow dx^D,\check{W}%
^{ab}\rightarrow \underline{W}^{ab}.  \label{19}
\end{equation}
Strictly speaking, one must perform a series expansion of the fields on the $%
x^D$ coordinate. The $x^D$-independent term in this expansion leads, with
the given identifications, to gravity in $d=D-1$ dimensions. The last term,
which does not contribute to the $D$-dimensional equations of motion, is the
Chern-Simons action for $\partial M_D$, i.e.,
\[
S_D^{\left( \text{G2}\right) }=\frac{\kappa _D}{\kappa _d}\int_{\partial M_D}%
{\cal L}_d^{\left( \text{CS}\right) }.
\]

The Chern-Simons term, $S_D^{\left( \text{G2}\right) }$, forces us to take
as $M_d$ the boundary of $M_D$, that is, $M_d=\partial M_D$. In this way the
freedom we initially had to pick any manifold out of the equivalence class
induced by $\sigma $ is lost, and we are left with a precise choice for $M_d$%
.

It is clear from the form of $S_D^{\left( \text{G1}\right) }$ that, in order
to obtain well-defined dynamics for the gravitational sector alone, one
needs to integrate the fields over the $x^D$ coordinate {\em before}
performing the identification process. This is due to the presence of the
extra $\left( d-2p\right) $ factor, which precludes $S_D^{\left( \text{G1}%
\right) }$ from leading to well-defined dynamics. Also, the zero mode\
hypothesis used in odd dimensions seems useless here, because we need some
kind of dependence on the $x^D$-coordinate to have a well-defined action for
gravity in $d$ dimensions. With this in mind, we shall take $M_d=\partial
M_D $ and parametrize the $x^D$-coordinate in such a way that it ranges
through $-\infty <x^D\leq 0$, with $x^D=0$ corresponding to the boundary of $%
M_D$. We shall additionally assume that the vielbein may be written as
\[
V^a=\exp \left( kx^D\right) V_0^a
\]
\[
V^D=dx^D,
\]
where $k$ is a real, positive constant with dimensions of [length]$^{-1}$
and $V_0^a$ and $W^{ab}$ are taken to be $x^D$-independent. Clearly, $V_0^a$
corresponds to $V_0^a=V^a\left( x^D=0\right) $.

With these assumptions, the action (\ref{SG1}) takes the form
\[
S_D^{\left( \text{G1}\right) }=\frac 1l\frac{\kappa _D}{\kappa _d}%
\int_{M_D}\sum_{p=0}^{\left[ d/2\right] }\alpha _p^{\left(
d\right) }\left( d-2p\right) \exp \left( k\left( d-2p\right)
x^D\right) \times
\]
\[
\times \varepsilon _{a_1\cdots a_d}{\cal R}^{a_1a_2}\cdots {\cal R}%
^{a_{2p-1}a_{2p}}V_0^{a_{2p+1}}\cdots V_0^{a_d}dx^D.
\]
Integration over $x^D$ from $x^D=-\infty $ to $x^D=0$ leads to
\[
S_D^{\left( \text{G1}\right) }=\frac 1{kl}\frac{\kappa _D}{\kappa _d}%
\int_{M_d}\sum_{p=0}^{\left[ d/2\right] }\alpha _p^{\left( d\right)
}\varepsilon _{a_1\cdots a_d}{\cal R}^{a_1a_2}\cdots
\]
\begin{equation}
\cdot \cdot \cdot {\cal R}^{a_{2p-1}a_{2p}}V_0^{a_{2p+1}}\cdots V_0^{a_d}.
\label{maga}
\end{equation}
The further addition of $S_D^{\left( \text{G1}\right) }$ and $S_D^{\left(
\text{G2}\right) }$ finally yields
\[
S_d^{\left( \text{red}\right) }=\kappa _d^{\left( \text{red}\right)
}\int_{M_d}\sum_{p=0}^{\left[ d/2\right] }\alpha _p^{\left( d\right)
}\varepsilon _{a_1\cdots a_d}{\cal R}^{a_1a_2}\cdots
\]
\begin{equation}
\cdot \cdot \cdot {\cal R}^{a_{2p-1}a_{2p}}V_0^{a_{2p+1}}\cdots V_0^{a_d},
\label{sdred}
\end{equation}
where
\begin{equation}
\kappa _d^{\left( \text{red}\right) }=\frac{\kappa _D}{\kappa _d}\left[
\frac 1{kl}+1\right] .
\end{equation}

It is now clear that the identifications (\ref{19}) lead to a well-defined
action for gravity in $d$ dimensions, since the coefficients in the action (%
\ref{sdred}) correspond to those given in (\ref{alfa}).

From eqs. $(56-59)$ of \cite{Sal03} we can see that (\ref{sdred}) can be
written as
\[
S_d^{\left( \text{red}\right) }=\kappa _d^{\left( \text{red}\right)
}\int_{M_d}{\cal L}_{CS}^{(2n-1)}
\]
where
\[
{\cal L}_{CS}^{(2n-1)}=\sum_{p=0}^{n-1}\frac{l^{-(2n-1-2p)}}{(2n-1-2p)}%
\varepsilon _{a_1\cdots a_d}R^{a_1a_2}\cdots
\]
\[
\cdot \cdot \cdot R^{a_{2p-1}a_{2p}}e^{a_{2p+1}}\cdots e^{a_d}.
\]

It is perhaps necessary to note that the procedure here developed
is also valid in the dimensional reduction from $D=odd$ to
$d=D-1.$ In this case

\begin{equation}
\left( D-2p\right) \alpha _p^{\left( D\right) }=\frac dl\frac{\kappa _D}{%
\kappa _d}\alpha _p^{\left( d\right) },
\end{equation}
so that the action (\ref{SD}) can be written as
\[
S_D=\frac dl\frac{\kappa _D}{\kappa _d}\int_{M_D}\sum_{p=0}^{d/2}\alpha
_p^{\left( d\right) }\varepsilon _{a_1\cdot \cdot \cdot a_d}{\cal R}%
^{a_1a_2}\cdot \cdot {\cal R}^{a_{2p-1}a_{2p}}V^{a_{2p+1}}\cdot \cdot
\]
\[
\cdot \cdot \cdot V^{a_d}V^D-2\int_{M_D}\sum_{p=1}^{d/2}p\alpha _p^{\left(
D\right) }\varepsilon _{a_1\cdots a_d}{\cal R}^{a_1a_2}\cdots
\]
\[
\cdot \cdot \cdot {\cal R}^{a_{2p-3}a_{2p-2}}{\cal R}^{a_{2p-1},D}V^{a_{2p}}%
\cdots V^{a_d}.
\]
From this expression it is apparent that it is possible to obtain an action
for $d$-dimensional gravity from the corresponding action in $D=d+1$
dimensions. Any identification in the spirit of (\ref{ww}) will do the job;
for example (\ref{19})$.$ Strictly speaking, one must perform a series
expansion of the fields on the $x^D$ coordinate. The $x^D$-independent term
in this expansion leads, with the given identifications, to gravity in $%
d=D-1 $ dimensions. This means that even-$d$ gravity obtained from its odd-$D
$ partner always posses an empty universe solution, no matter what choice is
made for the $M_d$-manifold inside the equivalence class induced by $\sigma $%
.

\section{Conclusions}

We have generalized the dimensional reduction mechanism of
Grignani-Nardelli \cite{Gri93} in a way that permits obtaining
Euler Chern Simons Gravity from Lovelock Born Infeld Gravity. The
failure of the procedure of Grignani-Nardelli in obtaining of the
appropriate coefficients that lead to the Chern-Simons has its
origin in the fact that both in even dimensions and in odd
dimensions the coefficients can be written uniquely as

\[
\alpha _p^{(d)}=\kappa \frac{l^{-(d-2p)}}{d-2p}\left(
\begin{array}{c}
n-1 \\
p%
\end{array}
\right)
\]
with $n=[(d+1)/2]$ which can split as
\[
\alpha _p^{(2n-1)}=\kappa \frac{l^{-(2n-1-2p)}}{2n-1-2p}\left(
\begin{array}{c}
n-1 \\
p%
\end{array}
\right) ,
\]
\[
\alpha _p^{(2n)}=\frac \kappa {2n}l^{-(2n-2p)}\left(
\begin{array}{c}
n \\
p%
\end{array}
\right) .
\]

The binomial coefficients that appear in both cases depend on $n$ and not on
the dimensionality of space-time. When we go from an even dimension $D$ to
an odd dimension $d=D-1$, $n$ remains constant. Thus the coefficients are
not reduced in a way that leads to a Chern Simons theory.

\begin{acknowledgments}
This work was supported in part by by grant FONDECYT through \#
1040624 (S) and \# 1010485. P.~S.,  F.~I. and E.~R. wish to thank
R.~Troncoso and J.~Zanelli for their warm hospitality in Valdivia
and for many enlightening discussions, as well as G. Rubilar for
his valuable insights on the mathematical techniques of
dimensional reduction.
\end{acknowledgments}

\end{document}